\documentclass[fleqn, sort&compress]{cas-dc}

\usepackage[numbers]{natbib}

\usepackage{amssymb}
\usepackage{amsmath}
\usepackage{tabularx}
\usepackage{graphicx}
\usepackage{dcolumn}
\usepackage{bm}
\usepackage[position=top]{subfig}
\usepackage{xcolor}
\usepackage[T1]{fontenc}
\usepackage{xspace}
\usepackage[normalem]{ulem}
\usepackage{todonotes}
\usepackage{soul}

\newcommand{\SI}{\textcolor{blue}{supplementary materials}\xspace}
\newcommand\redsout{\bgroup\markoverwith{\textcolor{red}{\rule[0.5ex]{10pt}{0.4pt}}}\ULon}
\newcommand{\uu}[1]{\ensuremath{\,\mathrm{#1}}}
\let\oldAA\AA
\renewcommand{\AA}{\ensuremath{\text{\oldAA}}}

\newcommand{\stkout}[1]{\ifmmode\text{\sout{\ensuremath{#1}}}\else\sout{#1}\fi}

\def\tsc#1{\csdef{#1}{\textsc{\lowercase{#1}}\xspace}}
\tsc{WGM}
\tsc{QE}
\tsc{EP}
\tsc{PMS}
\tsc{BEC}
\tsc{DE}

\begin{document}
\let\WriteBookmarks\relax
\shorttitle{Modelling echanical properties of amorphous phases}
\shortauthors{G.K. Nayak et~al.}

\title[mode = title]{Accurate prediction of structural and mechanical properties on amorphous materials enabled through machine-learning potentials: a case study of silicon nitride}

\author[1]{Ganesh Kumar Nayak}[type=editor,
                        orcid=0000-0002-6996-8998]

\fnmark[1]
\ead{nayak@mch.rwth-aachen.de}

\address[1]{Department of Materials Science, Montanuniversität Leoben, Franz-Josef-Strasse 18, 8700 Leoben, Austria}

\author[2]{Prashanth Srinivasan}
\address[2]{Department of Materials Design, Institute for Materials Science, University of Stuttgart, Pfaffenwaldring 55, 70569 Stuttgart, Germany}
\author[1]{Juraj Todt}
\author[1]{Rostislav Daniel}
\author[3]{Paolo Nicolini}
\address[3]{Institute of Physics (FZU), Czech
    Academy of Sciences, Na Slovance 2,
  18200 Prague, Czechia}

\author%
[1]
{David Holec}[orcid=0000-0002-3516-1061]
\ead{david.holec@unileoben.ac.at}
\ead[URL]{http://cms.unileoben.ac.at}

\fntext[fn1]{Current affiliation: Materials Chemistry, RWTH Aachen University, Kopernikusstraße 10, 52074 Aachen, Germany}

\begin{abstract}
    Amorphous silicon nitride (a-SiN) is a material which has found
    wide application due to its excellent mechanical and electrical
    properties.
    Despite the significant effort devoted in understanding how the
    microscopic structure influences the material performance, many
    aspects still remain elusive.
    If on the one hand \textit{ab initio} calculations respresent the
    technique of election to study such a system, they present severe
    limitations in terms of the size of the system that can be
    simulated.
    Such an aspect plays a determinant role, particularly when
    amorphous structure are to be investigated, as often results
    depend dramatically on the size of the system.
    Here, we overcome this limitation by training a machine-learning (ML)
    interatomic model to \textit{ab initio} data.
    We show that molecular dynamics simulations using the ML model on much larger
    systems can reproduce experimental measurements of elastic properties,
    including elastic isotropy.
    Our study demonstrates the broader impact of machine-learning
    potentials for predicting structural and mechanical properties, even
    for complex amorphous structures.
\end{abstract}




\maketitle

\section{Introduction}
\label{sec:intro_a-SiN}

Silicon nitride is a ceramic material of great technological interest
with diverse applications owing to its good mechanical and electrical
properties~\cite{Riley2004-ed, habraken1994silicon,katz1980high,
  liu1990structural, gomes1999tribological, hyuga2004influence,
  ishigaki1986friction, jones2001mechanical,
  cavaleiro2007nanostructured}.
Si$_3$N$_4$ can be synthesized using various processes such as
sputtering, chemical vapor deposition, and glow-discharge
decomposition~\cite{Milek1971-iu}.
Sintered Si$_3$N$_4$ components exhibit high density, high melting
temperature, low mechanical stress, high thermal strength, strong
resistance against thermal shock, and fracture toughness, and are used
in many engineering applications~\cite{klemm2010silicon,
  morgenthaler1994tailoring}.
For example, they are used for making engine components and cutting
tools due to their superior mechanical properties at high
temperatures~\cite{katz1980high, liu1990structural}.
SiN$_x$ in the nanocomposite TiN/SiN$_x$ acts as a protective coating
due to its high hardness and excellent
wear-resistance~\cite{gomes1999tribological, hyuga2004influence,
  ishigaki1986friction, jones2001mechanical}.
Incorporating Si$_3$N$_4$ as a second phase has been proposed to
improve the tribological performance of a material, while keeping
the other wear properties intact~\cite{jones2001mechanical,
  cavaleiro2007nanostructured}.

Additionally, Si$_3$N$_4$ in the amorphous form also has several
technological benefits.
Thin films of a-Si$_3$N$_4$ exhibit a high dielectric constant, a
high-energy barrier for impurity diffusion, high resistance against
radiation, and show oxidation resistance up to 1500\,$^{\circ}$C, making
them ideal candidates for several microelectronic
applications~\cite{gupta1988plasma, ma1998making, milek2013silicon} and as a gate dielectric in thin-film transistors~\cite{powell1981amorphous}.
Thick films of Si$_3$N$_4$ are promising candidates for non-linear
optical applications~\cite{ning2012strong, moss2013new}.
The above properties together with the bio-compatibility of a-SiN,
also make it an exceptional candidate for bearings in hip and knee
joint replacements~\cite{bal2012orthopedic}.

Several previous studies have focused on the TiN/SiN$_x$
nanocomposites reporting the structure and strength of the
interfaces~\cite{Alling2008-jm, zhang2009electronic, hao2006structure,
  ivashchenko2012comparative, zhang2010understanding}.
Yet, all these studies consider only crystalline structures of
stoichiometric Si$_3$N$_4$ and a few other specific stoichiometries of
SiN$_x$.
Understanding the structural and mechanical properties of the
amorphous stoichiometric silicon nitride (a-Si$_3$N$_4$) is relevant
due to its usefulness in nanocomposite structures.
In particular, analyzing the relationship between the structural and
mechanical properties of a-Si$_3$N$_4$ becomes essential to fine-tune
the fabrication of functional films.
The quantities explaining the structural characteristics, \textit{e.g.},
densities, bond distribution, \textit{etc}., of amorphous
Si$_3$N$_4$ have been studied with statistical approaches to capture the
effect of inherent local variability in the atomic
structure~\cite{Vedula2012-kf}.
The investigated structural properties are scalar quantities and
vary with different parameters, \textit{e.g.}, experimental
deposition condition~\cite{Vila2003-oe, Le2013-gm, Umesaki1992-qm,
  De_Brito_Mota1998-tw, kroll2001structure, Giacomazzi2009-fl}.
To a good degree they can be explained with appropriate semi-empirical
models, \textit{e.g.}, \citet{Milardovich2023-we} calculated RDF from a few empirical potentials with an excellent agreement.
However, an extension of their application toward the tensorial properties of the
materials, such as the elastic response, is not straightforward.
The mechanical properties of SiN$_x$ have been investigated by
experiments and modeling~\cite{Vedula2012-kf, Vila2003-oe,
  Le2013-gm, Zhou2023-pe, Khan2004-ob}.
However, this proves to be difficult to do fully at the \textit{ab initio} level, owing to the constraint of small system sizes, which has a particular impact on tensorial, \textit{e.g.}, mechanical properties.
In the wider context, the study of mechanical response w.r.t. the
effect of local atomic structure and the system size in amorphous
material is still lacking.
Computational techniques may provide a way to predict these properties
accurately and extensively.

The accuracy of the results depends on the computational method
employed.
\textit{Ab initio} electronic structure methods are extremely
accurate, yet limited to a few hundred atoms owing to their
computational cost.
A model limited to a few hundred atoms may not be large enough to
realistically model an amorphous phase, and the estimation
  of (some of) its properties may turn out to be quite
  inaccurate.
The elastic response of amorphous materials requires a more complex
description as it is not independent of the system size.
Hence, one must resort to larger-scale methods based on empirical
  potentials in order to obtain properties comparable to experiments.
The accuracy of such simulations depends on the interatomic potential
that defines the system's interactions.
Two empirical models that are available in the
literature~\cite{Umesaki1992-qm, Loong1995, Vedula2012-kf}
have been used to identify several structural properties of silicon
nitride~\cite{Umesaki1992-qm, Kalia1997-wy, Kalia1997-xt}.
However, neither of these models has been employed to study the
  elastic properties of a-Si$_3$N$_4$, and, given the fact that they
  consist of a bonded part, the modeling of formation and breaking of
  chemical bonds (\textit{e.g.} in a melt/quench simulation) remains
  questionable.
\citet{Vedula2012-kf} showed the distribution of bulk modulus of
stress-relaxed a-Si$_3$N$_4$ which varies as a function of densities
using a modified Born-Mayer-Huggins empirical force
field.
However, the variation of Young's modulus is still missing.
Owing to the complexity of the configurational space of amorphous
structures, it becomes difficult to develop accurate classical and reactive interatomic models.
A possible solution to capture the complexity is to train and use one
of the recently developed machine-learning interatomic potentials
(MLIPs) for atomistic simulations~\cite{Behler2007-uw, Bartok2010-mg, Bartok2017-nc, Shapeev2016-ew, Smith2017-nh}.
MLIPs can perform simulations with accuracy comparable to DFT at
a computational cost of orders of magnitude lower.
They are highly accurate and robust even for complex systems,
making them much more suitable than previous empirical models.
Moreover, most models often yield linear scaling behavior with the system size~\cite{bochkarev2022efficient}.

In this work, we train a moment tensor potential (MTP) -- a class
of MLIPs~\cite{Shapeev2016-ew} -- to model amorphous
Si$_3$N$_4$.
MTPs have been used to study several material systems ranging from
unaries~\cite{novoselov2019moment},
alloys~\cite{mortazavi2021first, tasnadi2021efficient}, and
multi-components~\cite{yin2021atomistic, grabowski2019ab,
  jafary2019applying} for the prediction of various properties,
such as diffusion, mechanical properties, vibrational free energies,
dislocation mobility, magnetism.
Nevertheless, they have not been trained to model an amorphous
structure yet. Here, we train an MTP and simulate the structural and
mechanical properties of a-Si$_3$N$_4$.
To validate the prediction of both the trained MTP and \textit{ab
  initio} calculations, we directly compare the outcomes to
experimental results.

\section{Methods and model construction}\label{sec:methandcompdetails_a-SiN}

At the beginning, \textit{ab initio} molecular dynamics (AIMD)
simulations were performed to determine the appropriate mass
density and the corresponding system size to be used in the
subsequent steps.
The training data for the MTP were then generated by means of
density-functional theory (DFT) calculations.
The MTPs were eventually used to perform classical molecular dynamics
(MD) simulations to predict the elastic constants.

\subsection{Ab initio calculations}
\label{subsec:ab-init_a-SiN}

\subsubsection{Prediction of mass density}
\label{subsubsec:equ_mass-density}

The procedure followed to obtain the size-independent mass
density was based on the `melt-quench'
approach~\cite{Aykol2018-ve, Aykol2018-lf}.
We started with a liquid and progressively lowered the temperature,
thereby 'freezing into' an amorphous structure.
The generation of the initial structure for the melt/quench
process was done so that random positions were assigned to N
and Si atoms preventing atomic overlap by enforcing a minimum
  distance of 2\,\AA\,between neighbors.
We did not consider the melting and quenching of crystalline structure
to avoid long simulations necessary to achieve the total melting
of the samples.
The workflow was implemented using the pymatgen software
package~\cite{Ong2013-iv} as follows:
(i) atoms were randomly distributed in cubic simulation cells of
different sizes; a total of 10 starting structures was generated
with system sizes ranging between 56 and 182 atoms, with a
volume 15\% larger than for crystalline Si$_3$N$_4$ (lattice
parameters were taken from Ref.~\cite{file1996database}),
(ii) the structures were melted at 2500\,K (the melting
the temperature of Si$_3$N$_4$ is $\approx2150$\,K~\cite{Hoffmann1995-hz}) and 0\,kB pressure for 5000 timesteps and subsequently quenched to
0\,K with $NPT$ ensemble using AIMD at a cooling rate of $0.5\times10^{-15}$\,K/s,
(iii) from both the melting and quenching processes, six structures
(for each system size) were chosen every 500 timesteps of AIMD run, after removing the initial 1500 steps of thermalization (a structural analysis confirmed that the
sampled structures were statistically independent),
(iv) each of the structures was relaxed, allowing to vary ionic
positions, cell shape, and volume, such that the stress
component on the box was reduced to nearly zero. The thermostat, barostat, the relaxation scheme, and termination criteria used in this protocol are discussed in Sec.~\ref{subsubsec:comp_a-SiN}.
Following this procedure allowed us (i) to identify an appropriate
  system size representative enough of the real material, with
  reasonable computational cost, (ii) to estimate the 
  density of the material, and (iii) to predict accurately
  structures in terms of box size and short-range order interactions
for further estimation of properties for comparison with MTP.

\subsubsection{Generation of training set for MTP}
\label{subsubsec:mtp_data_gen}

The data for the training set (TS) to be used for the
parameterization of the MTP was taken from AIMD calculations
performed on 224-atom simulation boxes (see
Fig.~\ref{fig:density} and related discussion in
Sec.~\ref{sec:structural_a-SiN}).
This particular system size was motivated by having a statistically robust distribution of local environments in our training datasets for the amorphous material.
The training data must contain box volumes of sufficient size spans to
account for thermal expansion, preferably up to the liquid phase
(relevant when the melting state is achieved in the melt-quench
process) or up to the properties of interest. 
We aimed to choose the volume for the optimal TS, which spans a large
area of the relevant phase space.
The argument was that to have 
sufficient volume variation to achieve different densities (in g/cm$^2$) in the TS to train MTP w.r.t. more or less densely packed neighborhoods (discussion in Sec.~\ref{sec:structural_a-SiN}). 
Therefore, we applied positive and negative strains, maximum up to
$\pm$4\% given in one lattice orientation with reference to
predicted mass density (see Fig.~\ref{fig:density}). 
Hence, the volume variations related to the mass densities 2.92, 3.05,
3.18 (average), 3.34, and 3.50 g/cm$^2$ was used. 
The starting configurations for these densities were achieved from the randomly filled boxes of corresponding sizes.
For the actual TS generation, AIMD simulations of 1500 steps each were performed in the $NVT$ ensemble at these five
volumes at a constant temperature of 1800\,K. 
The total size of the training dataset consisted of 5000 structures, excluding the first
500 initial AIMD steps for each simulation corresponding to the
initial equilibration processes.

\subsubsection{Structural and elastic properties}
\label{subsubsec:elastic_a-SiN}

To compute the radial distribution function (RDF) for further MTP validation, a separate AIMD simulation with 112 atoms was performed with the $NVT$ ensemble. 
The starting configurations were achieved from the `melt-quench' of atoms randomly distributed in box, by fixing the volume to achieve predicted (mean) mass density equivalent to MTP prediction and followed by the equilibration at room temperature (to compare with MTP and experiment), with 5000 steps.

The elastic response of a-Si$_3$N$_4$ is calculated with a
stress-strain approach~\cite{Yu2010-jx, Zhou2013-iy}, which generally
produces a 4th-order tensor of elastic constants, which can
be represented with a $6 \times 6$ matrix ($C_{ij}$ in Voigt's
  notation~\cite{Nye1985-vz}).
To achieve this, the reference structure(s) was(were) relaxed, allowing to vary ionic positions, cell shape, and volume until an internal pressure of 0\,bar is realized.  Given the original matrix, the new lattice vectors were achieved by applying a set of strains up to $\pm2\%$ (6 positive and 6 negative) by modifying the unit cell lattice vectors. These strain-induced structures were relaxed to compute stress, allowing only varying ionic positions.
The directional elastic response was further analyzed using the ELATE package~\cite{Gaillac2016-ss}. 
To quantify the anisotropy, we employed the tensorial anisotropy index~\cite{Sokolowski2018-mj}:
\begin{equation}
  A^T = \frac{2 \left( C_{44} + C_{55} + C_{66} \right)} {\left(
    C_{11} + C_{22} + C_{33} \right) - \left( C_{12} + C_{13} + C_{23}
    \right)}.
  \label{eq:isotropic_response}
\end{equation}
The closer the value of $A^T$ is to $1$, the more
  isotropic is the elastic tensor.

\subsubsection{Computational details of ab initio calculations}
\label{subsubsec:comp_a-SiN}

All AIMD and DFT simulations were done using the Vienna \textit{Ab
Initio} Simulation Package (VASP)~\cite{Kresse1996-dg,
Kresse1996-mc}.
We used projector augmented-wave (PAW) potentials~\cite{Blochl1994-ub}
along with the Perdew, Burke, and Ernzerhof (PBE) parametrization
of the generalized gradient approximation (GGA)~\cite{Perdew1996-jm}. 
A plane-wave cutoff of 500\,eV and Methfessel–Paxton~\cite{methfessel1989high} smearing of 0.2\,eV was also used in all calculations.

The AIMD calculations for the density convergence and generating training sets for MTP used a single $\Gamma$-centered $k$-point mesh to reduce the computational cost.
In the $NPT$-simulations of mass density prediction (see Sec.~\ref{subsubsec:equ_mass-density}), the AIMD was performed using a time-step of 1\,fs with the Langevin thermostat and the Parinello-Rahman barostat. 
To generate the TS and to compute structural properties, we performed AIMD with the Nose-Hoover thermostat using a time-step of 3\,fs and 2\,fs, respectively.

The DFT relaxation of the structures, after AIMD, was done using Brillouin zone sampling with $3\times3\times3$ Monkhorst–Pack~\cite{monkhorst1976special} $k$-point mesh, and a convergence criterion of $10^{-6}$\,eV (per supercell) for the electronic self-consistency cycles, and that of ionic relaxations upto $10^{-3}$\,eV (per supercell). 
The DFT calculation of elastic constants (Sec.~\ref{subsubsec:elastic_a-SiN}) employed a
$\Gamma$-centered $6\times6\times6$ Monkhorst–Pack $k$-point scheme.

\subsection{MTP training}\label{sec:mtp} 

\begin{figure*}
    \centering
    \includegraphics[width=14cm]{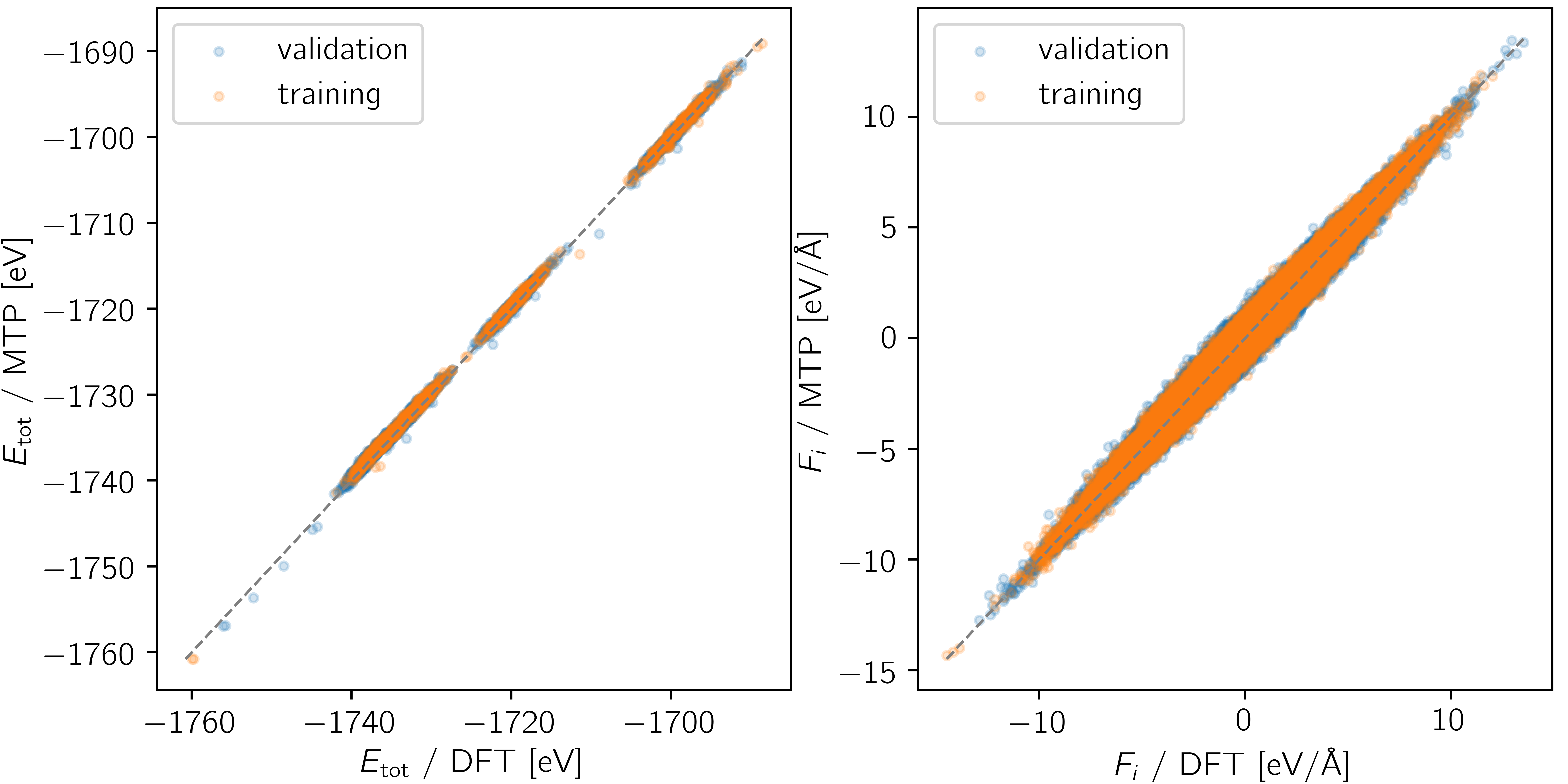}
    \caption{Validation of the fitted MTP.}
    \label{fig:MTP_validation}
\end{figure*}

To perform MD calculations on a-Si$_3$N$_4$, we trained an MTP to the DFT data.
We fitted the MLIP using the MTP-based code mlip-v2~\cite{Shapeev2016-ew}. Related to the highest degree of polynomial-like basis functions used in the analytic description of the MTP, an initial MTP of level-24g is used to define its functional form. 
The cut-off radius of $7\,\uu{\AA}$ was set in the initial MTP.
The first MLIP was derived by fitting the initial MTP to 167 structures in the TS using the Broyden-Fletcher-Goldfarb-Shanno method~\cite{Fletcher2000-ga} with 1000 iterations and fitting weights of 1.0, 0.01, and 0.001 for the total energy, atomic forces, and stress in the loss function, respectively.
In the next step of fitting, the first MLIP was used as the initial MTP, and so on. 
This process was repeated four times by amending the TS with structures identified by the ``select-add'' method as implemented in the mlip-v2.
In this process, the final training dataset contained 1527 structures out of the 5000 structural snapshots calculated by AIMD. We used the remaining 3473 structures as a validation set. 
The energies (total energy, 1 value per structure) and forces (3 values per atom, 224 atoms per structure) calculated by the MTP are plotted against the DFT values in Fig.~\ref{fig:MTP_validation}.
There is no visible difference between the behavior of the training and validation data.
The final root-mean-square errors (RMSEs) in the energy and forces of the final
fitted MTP are 4.5\,meV/atom and 0.27\,eV/\AA,
respectively, for the training set, which further decrease to 2.0\,meV/atom and 0.22\,eV/\AA, respectively, for the complete dataset.

\subsection{Classical MD simulations with MTP}\label{sec:classical}
To prove the size-dependence of structural and elastic properties of amorphous SiN, we performed
MD simulations using the large-scale Atomic/Molecular Massively Parallel
Simulator (LAMMPS)~\cite{Plimpton1995-tf} code together with the fitted MTP potential.

In the first place, two structures with 112 and 7000 atoms were generated with an analogous methodology to the AIMD as discussed in the Sec.~\ref{subsubsec:equ_mass-density}. 
However, in contrast to AIMD, after randomly distributing atoms in a box with a size corresponding to the DFT-estimated mass density, the systems were first annealed (i.e. temperature was gradually ramped) up to 2500\,K at 0\,bar pressure over 5\,ps, followed by 5\,ps melting and quenched back to 0\,K (\textit{i.e.}, `melt-quench') in $NPT$-ensemble of LAMMPS simulations with the fitted MTP. 
The last two steps are identical to those performed in the AIMD melt-quench procedure.
To be consistent with AIMD, we considered the heating and cooling rates of $10^{-15}$\,K/s but used a smaller timestep of 0.25\,fs; by running the simulations for steps of 20000 and 10000, respectively, we achieved the same total times as in AIMD.
To analyze the impact of cooling rates on the defects, for the system with 7000 atoms, we also consider the simulation at cooling rates of $10^{-13}$ and $10^{-14}$\,K/s achieved by changing the number of steps to 1000000 and 100000, respectively, while keeping the timestep constant.
The resulting structure obtained from this $NPT$ simulation was structurally fully relaxed at 0\,K and subsequently used to calculate the elastic constants for comparison with the corresponding DFT results. 
These 0\,K the elastic constants were computed using the stress-strain method (described in Sec.~\ref{subsubsec:elastic_a-SiN}), where the LAMMPS code was used to apply the deformation up to $\pm2\%$.

The elastic constants at finite temperatures were calculated using averaged stress and strain extracted from the distortions by employing the $NVT$ simulations and averaging the stress tensor with a fully equilibrated deformed cell. 
We used 1000 steps for each deformation for the equilibration followed by 300 steps for the production run with a timestep of 0.1\,fs.
The small timestep was chosen to minimize stress fluctuations and related statistical errors.

The temperature-dependent elastic constants were calculated for temperatures ranging from $\approx300$\,K to $\approx1600$\,K with steps of $\approx150$\,K. The initial configurations for the calculation of elastic constant for these temperatures were achieved by the equilibration simulation with $NVT$ ensemble initialized from structural snapshots from the `quench-melt' NPT run. In all our $NPT$ and $NVT$ simulations, we used the Nose-Hoover thermostat and Nose-Hoover barostat for $NPT$ ensembles. All the elastic tensors are evaluated using the too ELATE: elastic tensor analysis~\cite{Gaillac2016-ss}. 
Elastic properties predicted by our MTP potential were further validated by simulations using a recently published GAP potential \cite{Milardovich2023-we} with the same MD simulation protocols.

\subsection{Experimental details}
The a-SiN films were sputter deposited with a thickness of 1--3\,$\mu$m to measure the composition, density, radial distribution function (RDF), and mechanical properties. The deposition was done on a single-crystalline Si substrate at 200$^\circ$C.
The deposition was accomplished in Ar/N$_2$ plasma discharge, where the partial pressure of nitrogen has varied from $0.95\times10^{-3}$ to $1.55\times10^{-3}$ mbar, while that of Ar partial pressure ranges from $3.1\times10^{-3}$ to $2.51\times10^{-3}$ mbar.
The composition measurement using energy dispersive X-ray analysis (EDX) revealed that the ratio of Si/N is approximately 3/4.
We note this could be potentially somewhat inaccurate, as EDX can yield up to 5\,\%at. errors. 
The samples have, however, a simulation-equivalent composition and, therefore, their results are discussed in this paper together with literature data.
All the measurements (mass density, radial distribution function (RDF), structure factor ($S(Q)$), nanoindentation) were performed at room temperature, \textit{i.e.}, 20--25$^\circ$C.

Experimental RDF was obtained using synchrotron wide-angle X-ray scattering (WAXS) at the high-energy materials science beamline (HEMS) of PETRA III at DESY in Hamburg, Germany. The sample was probed with a monochromatic beam of $100\times10\,\mu$m$^2$ (horizontal$\times$vertical) and wavelength of 0.1695\,$\uu{\AA}$, with scattered intensities recorded on a Perkin Elmer XRD1621 2D flat panel detector placed 1065.4 mm downstream of the sample. Integration around the direct beam was performed after careful
background signal subtraction and masking of shadowed detector regions. Subsequent RDF calculation was performed with LiquidDiffract software~\cite{Heinen2022-er} using a mass density of 2.7\,g/cm$^3$ experimentally determined by X-ray reflectometry (XRR) using a laboratory device and GenX software~\cite{Glavic2022-jt}, low-$Q$ and high-$Q$ cutoffs (0.6 and 8.65\,$\uu{\AA}^{-1}$), a low-$r$ cutoff at 1.1\,$\uu{\AA}^{-1}$ and final Lorch-filtering. We have used two different indenters to test the mechanical properties: the UMIS nanoindenter and the Hysitron T950 Tribometer to cross-validate the results.

\section{Results and Discussion}\label{sec:results_discussion} 

We first report on the \textit{ab initio} prediction of
structural properties of a-Si$_3$N$_4$.
Then, we present the properties predicted by the MTP-based MD simulations and benchmark them against the \textit{ab initio} results. 
Both methods are also employed to predict the elastic properties
of amorphous Si$_3$N$_4$, and the results are used to discuss the
impact of the model size on the elastic response of the amorphous
phase. 
Finally, the theoretical predictions are validated against experimental measurements.

\subsection{Structural properties}\label{sec:structural_a-SiN} 

\begin{figure}[tb]
  \centering
  \includegraphics[width=\linewidth]{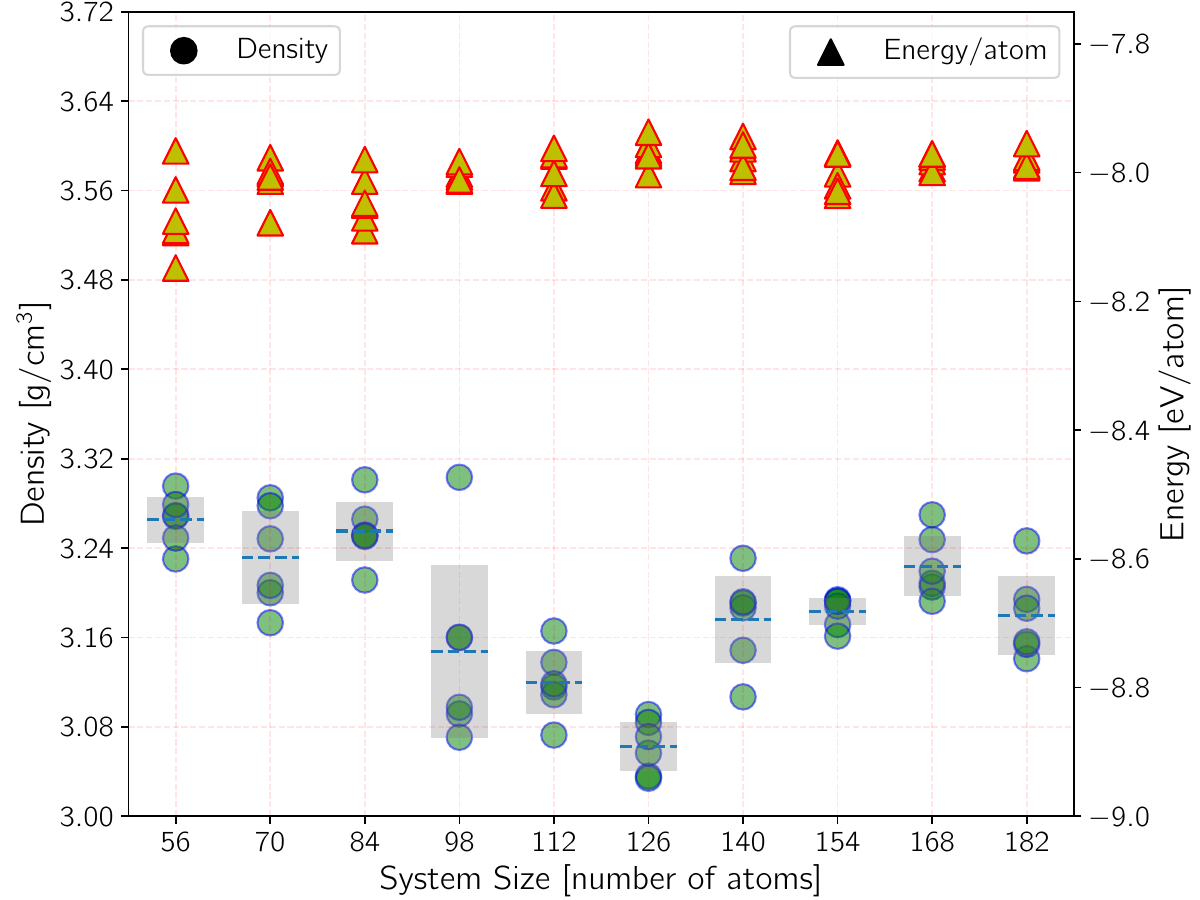}
  \caption{Mass density and total energy data from the AIMD
    `melt-quench' method followed by a DFT relaxation. The calculations have been done for six structural models for
    each system size with the stoichiometric composition of amorphous
    Si$_3$N$_4$.}
  \label{fig:density}
\end{figure}
 
Fig.~\ref{fig:density} presents the predicted 
mass density and total energy per atom by the
procedure described in Sec.~\ref{subsubsec:equ_mass-density}; for each
system size, values for all six tested structures are
shown. 
The values of the mass density of all the systems are scattered, and the six structures do not
converge to a narrow range of values. 
The average mass densities scatter between $3.08$\,g/cm$^3$ (126 atoms) and $3.25$\,g/cm$^3$ (84 atoms). 
The minimum and maximum values of all the mass densities are 3.03\,g/cm$^3$ and 3.30\,g/cm$^3$, respectively.
The mean across all system sizes, supposedly representing an ensemble over different local environments in the amorphous material, is $3.18\pm0.06$\,g/cm$^3$; this value will be considered further. 
The 112- and 154-atom systems have the least standard deviations.
Considering the computational DFT costs, which scale with a number of atoms cubed, the system with 112 atoms was chosen for further elasticity and validation calculations with DFT and AIMD.
This system size is also compatible with previously reported DFT and MD data (\textit{e.g.}, in Ref.~\cite{Vedula2012-kf}). 
As for the training dataset, to
be on the safe side regarding the representativeness of the ensemble of local environments, we chose the system size with 224 atoms. 
The fitting procedure is detailed in Sec.~\ref{sec:mtp}.

We now move towards analyzing structural properties from models calculated using classical MD with our fitted MTP.
This will allow us to discuss also their size-dependence beyond what DFT is capable of handling.
The structure was achieved by annealing, equilibrating at 2500\,K, and quenching back to 0\,K of an initial configuration with density of 3.18\,g/cm$^3$ obtained from the DFT calculations. The quenching rate considered was $10^{15}$\,K/s.
Results of analogous simulations with other heating and quenching rates are reported in \SI.
With the structure, we performed the structural analysis regarding the radial distribution
function, structure factor, and bond distribution to verify the
MTP properties against \textit{ab initio} MD.
The RDF and $S(Q)$ are calculated using the diffraction module implemented in the freud library~\cite{freud2020}.

\begin{figure}[tb]
  \centering
  \includegraphics[width=\linewidth]{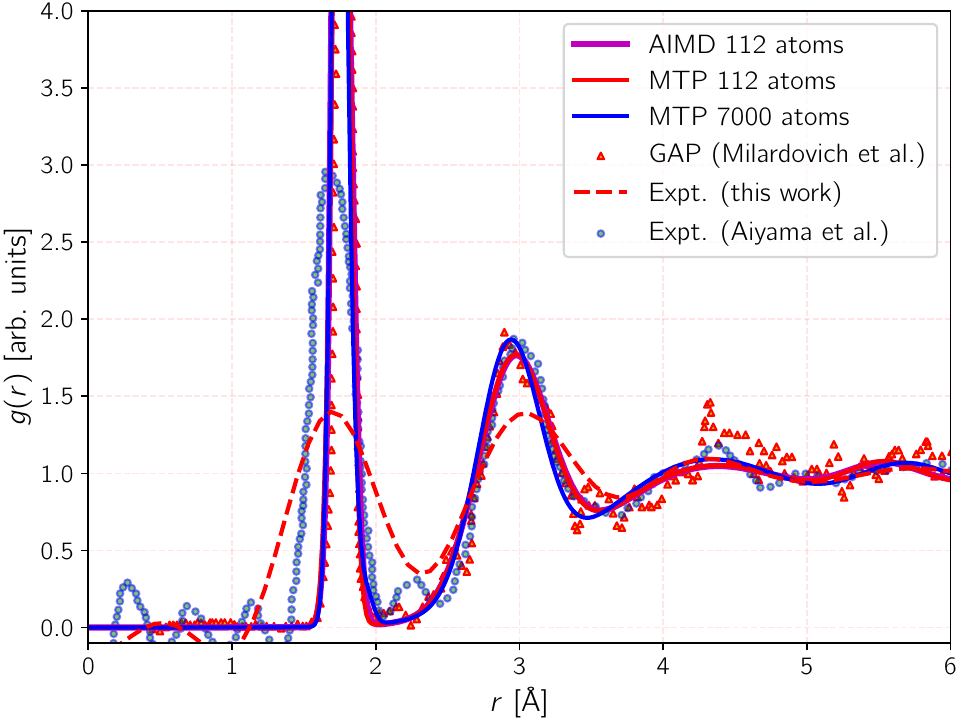}
  \caption{The total radial distribution function of a-Si$_3$N$_4$
    from the MTP-based molecular dynamics (with 112 and
    7000 atoms) is compared with the \textit{ab initio} model
    (112 atoms), and measurement. Our calculated RDF agrees well with
    the previous calculation by
    ML-GAP (224 atoms)~\cite{Milardovich2023-we}.
    }
  \label{fig:rdf}
\end{figure}

Importantly, our calculations suggest that the resulting structures, as measured by RDFs, are independent of quenching rates  (see \SI). 
Fig.~\ref{fig:rdf} shows RDFs for amorphous Si$_3$N$_4$ from various simulation models (all using the quenching rate of $10^{15}$\,K/s) and measured experimentally by XRR. 
The major features of the RDFs from all models, e.g. from MTP and from AIMD, are consistent. 
More importantly, the calculated RDFs agree reasonably well with
the experimentally measured ones, although the present experimental curve suffers from a significantly larger broadening of the first peak at around 1.6\,\AA.
This peak corresponds to the Si-N bonds, suggesting that there is a much larger variability of this bond length in our sample as compared with the theory of previous experimental measurement. 
Some minor deviations between the experiment and theory may, at least in part, be because
the reciprocal lattice vector dependence of the X-ray atomic
scattering factors (only atomic numbers were considered) was neglected
during the calculation of the theoretical RDF.

\begin{figure}[tb]
  \centering
  \includegraphics[width=\linewidth]{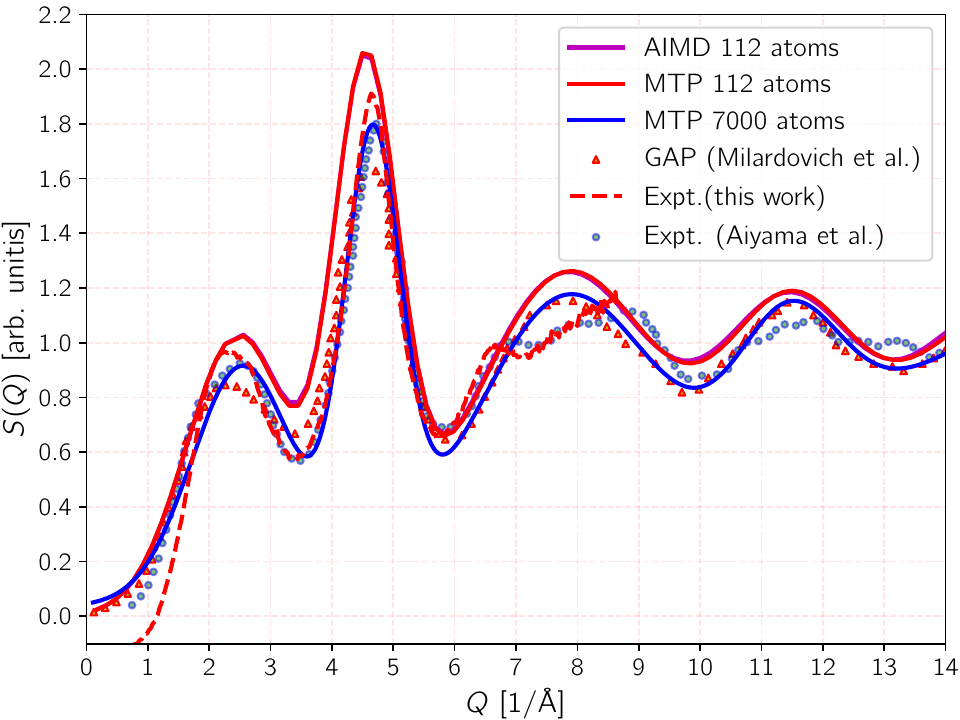}
  \caption{Computed structure factor $S(Q)$ by \textit{ab initio} and
    MTP, and obtained from the X-ray diffraction experiment for a
    well-annealed sample of a-Si$_3$N$_4$. Our calculated $S(Q)$ well
    agrees with the previously calculated
    ML-GAP~\cite{Milardovich2023-we}.
    }
  \label{fig:structure_factor}
\end{figure}

Fig.~\ref{fig:structure_factor} shows the computed structure factors,
$S(Q)$, which also compares well with the experimental
diffraction structure factor. 
The MTP curves for 112 atoms and 7000 atoms basically overlap, while small differences w.r.t. AIMD can be seen regarding the lowest $Q$ peak position, as well as amplitudes of the first two peaks.
Nevertheless, this satisfactory agreement between MTP and the \textit{ab initio} data confirms the ability of the fitted MTP model to reproduce the \textit{ab initio} model. 
Worth noting is also an excellent agreement of the $S(Q)$ against the
previously calculated GAP values~\cite{Milardovich2023-we}. 
However, for the measured $S(Q)$, there is a slight shift of the first
peak to the left ($\sim1.5-2.5\uu{\AA}^{-1}$). 
The deviation among the $S(Q)$s could be due to the finite-size
effects. 
In finite-size systems, $S(Q)$ is affected by the errors in the
Fourier transform by producing relative errors in the asymptotic
region of the $N$-particle function
$g_\text{N}(r)$~\cite{Salacuse1996-el}. 
Consequently, the first MTP-based calculated $S(Q)$ peak has slightly
shifted to the left compared to \textit{ab initio}. 
On the contrary, in our measured $S(Q)$ there is a slight mismatch at
the lower values ($\sim0.0$--$1.0\uu{\AA}^{-1}$); it might be because,
the interactions are most likely averaged and weighted by scattering
power; therefore, the interactions are mostly the Si-Si and Si-N bond
lengths and the N-N bonds to a lesser degree. 
The missing N-N bonds comprise the $S(Q)$ lower part.

It is also worth noting, that since RDF and $S(q)$ are related by Fourier transformation, they essentially convey the same information.

\subsection{Bonding and elastic properties}
\label{subsec:bonding-elastic_a-SiN}

\begin{figure}[tb]
    \centering
    \subfloat[\centering]{\label{fig:bond_length}
      {\includegraphics[width=\linewidth]{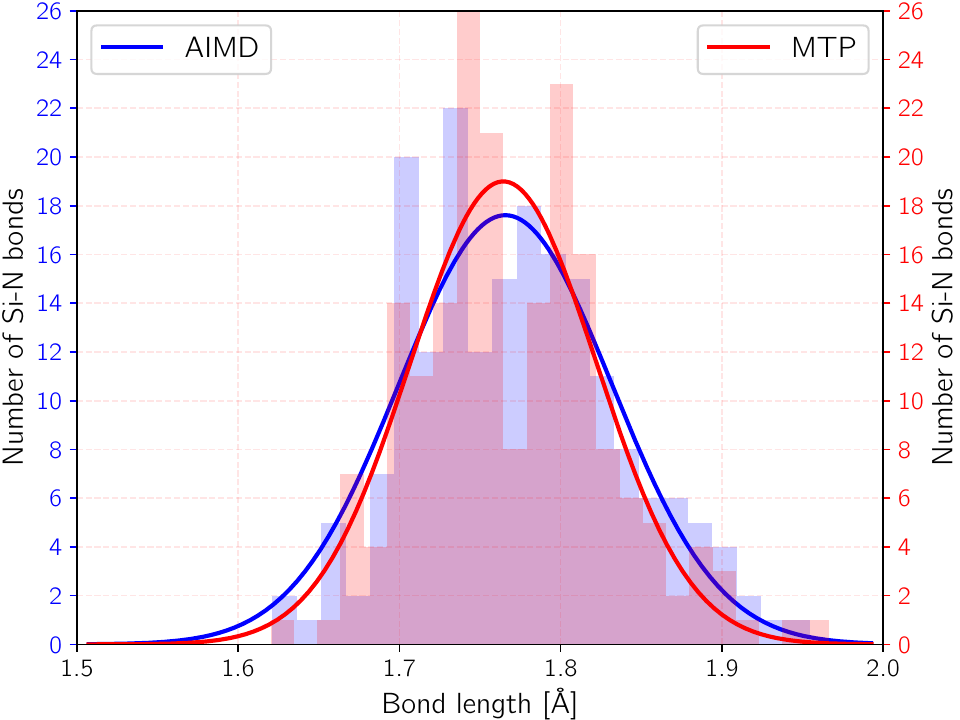}}}%
    \qquad
    \subfloat[\centering]{\label{fig:bond_angle}
      {\includegraphics[width=\linewidth]{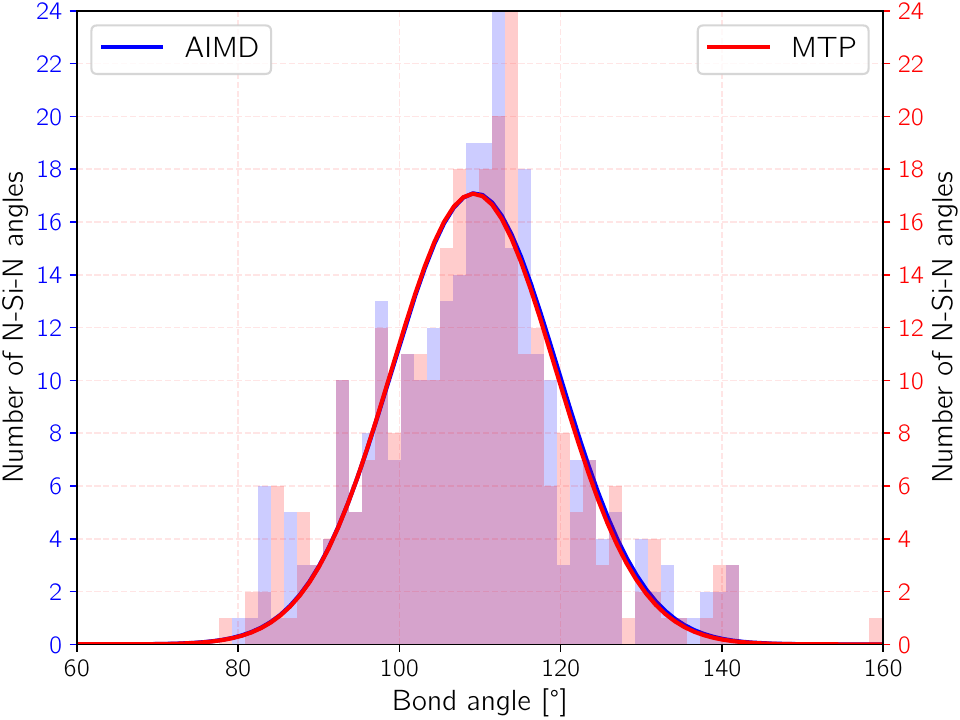}}}
    \caption{A comparison of structural models with 112 atoms generated by MTP and \textit{ab initio}, in terms of (a)~bond-length and (b)~bond angle distributions. The distributions
      are fitted with Gaussian profiles as a guide for the eye.}
    \label{fig:bond_distribution}
\end{figure}

To allow for a quantitative comparison of the MTP and \textit{ab
initio} models of 112 atoms, we fitted the bond length and bond angle histograms
with Gaussian distributions, as shown in
Figs.~\ref{fig:bond_length} and~\ref{fig:bond_angle}, respectively.
We note that these are used as visual aids and do not necessarily represent a physics-informed distribution of these quantities.
The average MTP bond length of $(1.780 \pm 0.093)\,\uu{\AA}$ well
reproduces the \textit{ab initio} bond length of $(1.773 \pm
0.093)\,\uu{\AA}$.
The two Gaussian distributions well agree, up to a small displacement, related to the slight disagreement of the two mean values above.

Similarly, good agreement is also obtained for the bond angles: MTP
yields $(109.22\pm10.23)^{\circ}$ and \textit{ab initio}
$(109.30\pm10.34)^{\circ}$.
It is worth noting the smaller standard deviation of the MTP model for
both quantities, likely related to more equilibration steps, despite the quenching rate of $10^{15}$\,K/s for both systems.
The bond angles are distributed around the ideal tetrahedral value of $109.5^{\circ}$ in a-Si$_3$N$_4$, Si being the center of the tetrahedron~\cite{Roizin2007-ty, Robertson1991-ko}.
A comparison of \textit{ab initio} model of 112 atoms vs. MTP 7000 atoms is shown in (see \SI).

We further analyzed the actual local coordination, and thereby
quantified defects with a deviation from the tetrahedron geometry.
The \textit{ab initio} simulated structure consists of 95\% of the
silicon atoms with a tetrahedral geometry, with 3- and 5-folded
bonded atoms of 2\% each. 
The MTP simulated structure almost mimics the \textit{ab initio} result,
having 92.8\% of the tetrahedron, 1.9\% of the 3-folded, and 5.3\% of
the 5-folded bond. 
These are results from the structure obtained with a quenching rate of
$10^{13}$ K/s. 
Slightly larger amounts of defects are obtained for faster quenching
rates (cf. Tab.~\ref{tab:stru_defects_quench}). 
However, we have realized from the RDF and $S(Q)$ that the quenching
rate (and thus the small differences in defect content) has little
impact on the structural changes (see \SI). 
Despite this little impact on the structural properties, a slow
quenching rate consistently yields models closer to defect-free structures
(\textit{i.e.}, with only 4-coordinated Si atoms).

\begin{table}[tb]
  \centering
  \begin{tabular}{c|c c c}
    \hline\hline
    & \multicolumn{3}{c}{Quenching Rate (K/S)} \\
    Coordination &  $10^{13}$ & $10^{14}$ & $10^{15}$\\
    \hline
    3-folded     & 1.9\% & 1.9\% & 2.7\% \\
    4-folded     & 92.8\% & 91.6\% & 90\% \\
    5-folded     & 5.3\% & 6.5\% & 7.3\% \\
    \hline\hline
  \end{tabular}
  \caption{Coordination of Si atoms in MTP models obtained with
    different quenching rates.}
  \label{tab:stru_defects_quench}
\end{table}

\begin{figure*}
  \centering
  \subfloat[\centering \textit{ab initio}: 112 atoms]{\label{fig:youngs_dft}
    {\includegraphics[width=0.5\linewidth]{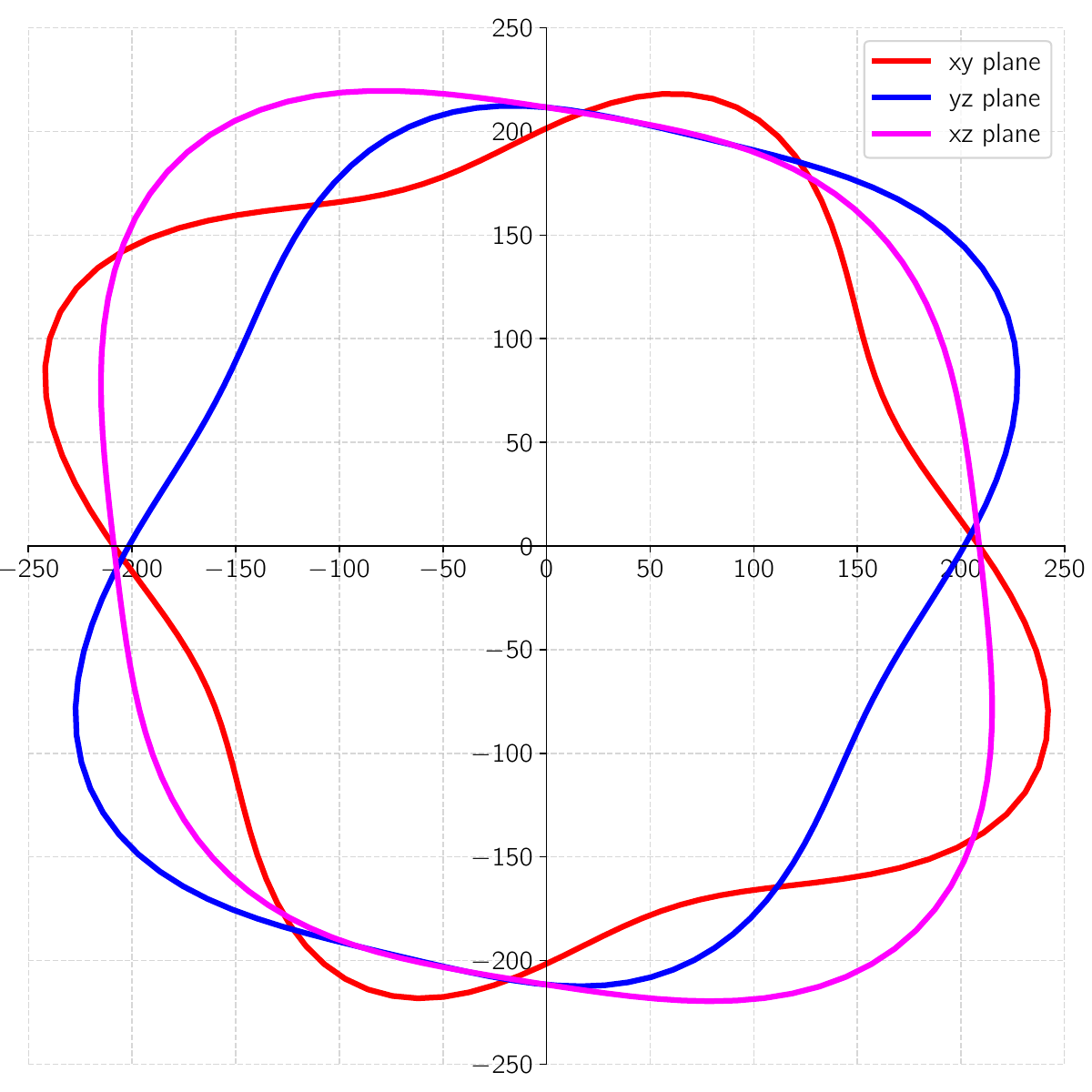}}}%
  \subfloat[\centering MTP: 112 atoms]{\label{fig:youngs_mlip_122}
    {\includegraphics[width=0.5\linewidth]{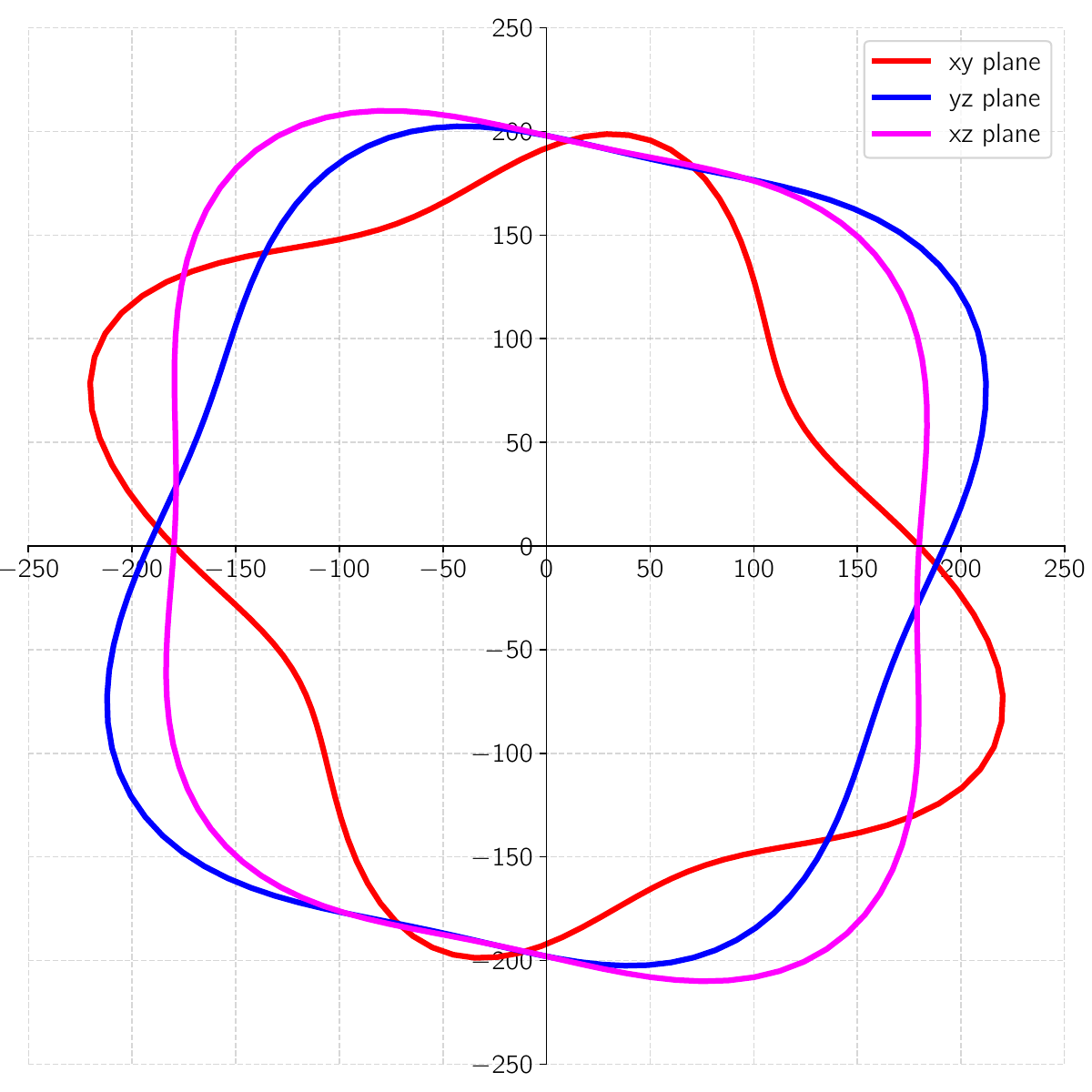}}}%
  \qquad
  \subfloat[\centering MTP:7000 atoms]{\label{fig:youngs_mlip}
    {\includegraphics[width=0.5\linewidth]{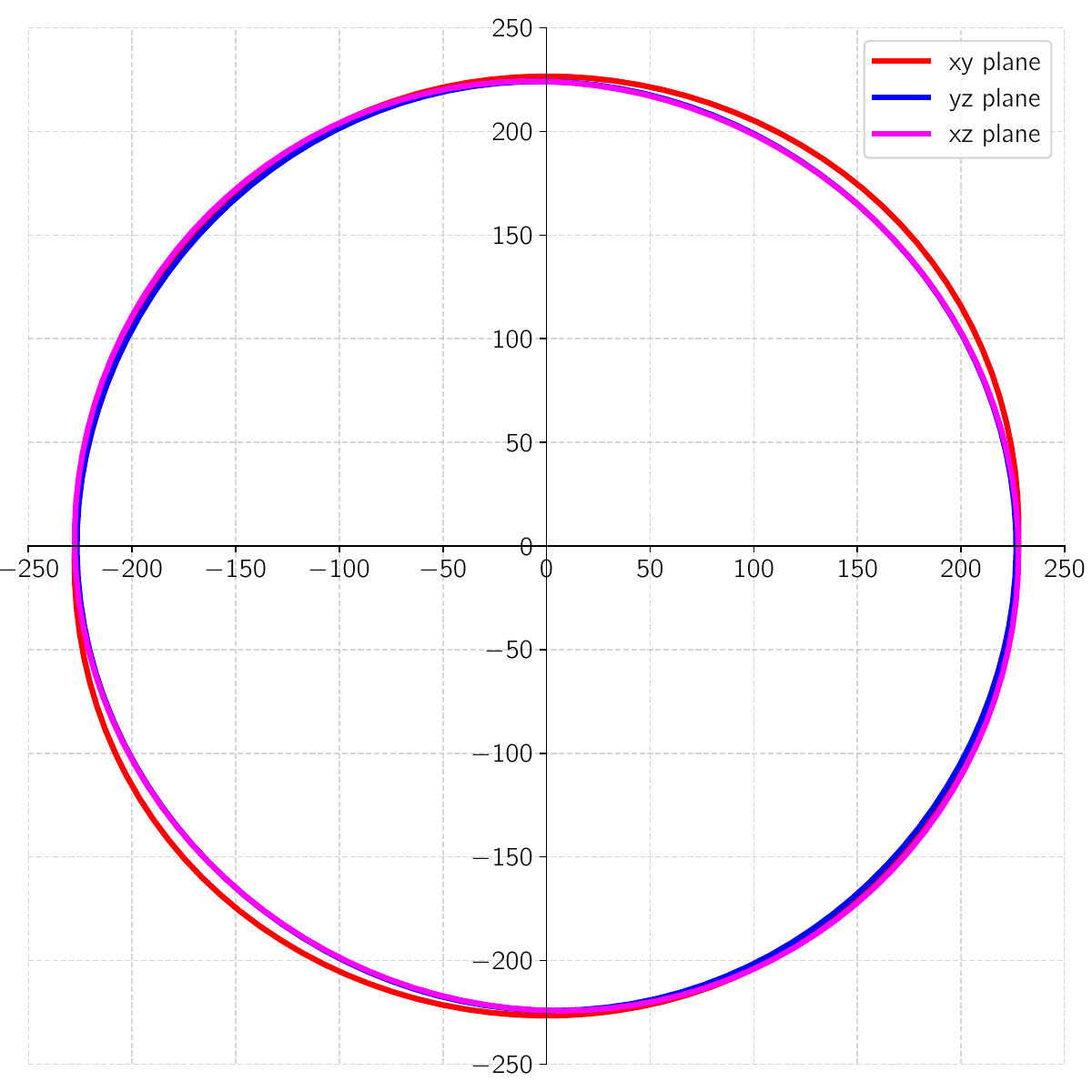}}}%
  \subfloat[\centering GAP:7000 atoms]{\label{fig:youngs_gap}
    {\includegraphics[width=0.5\linewidth]{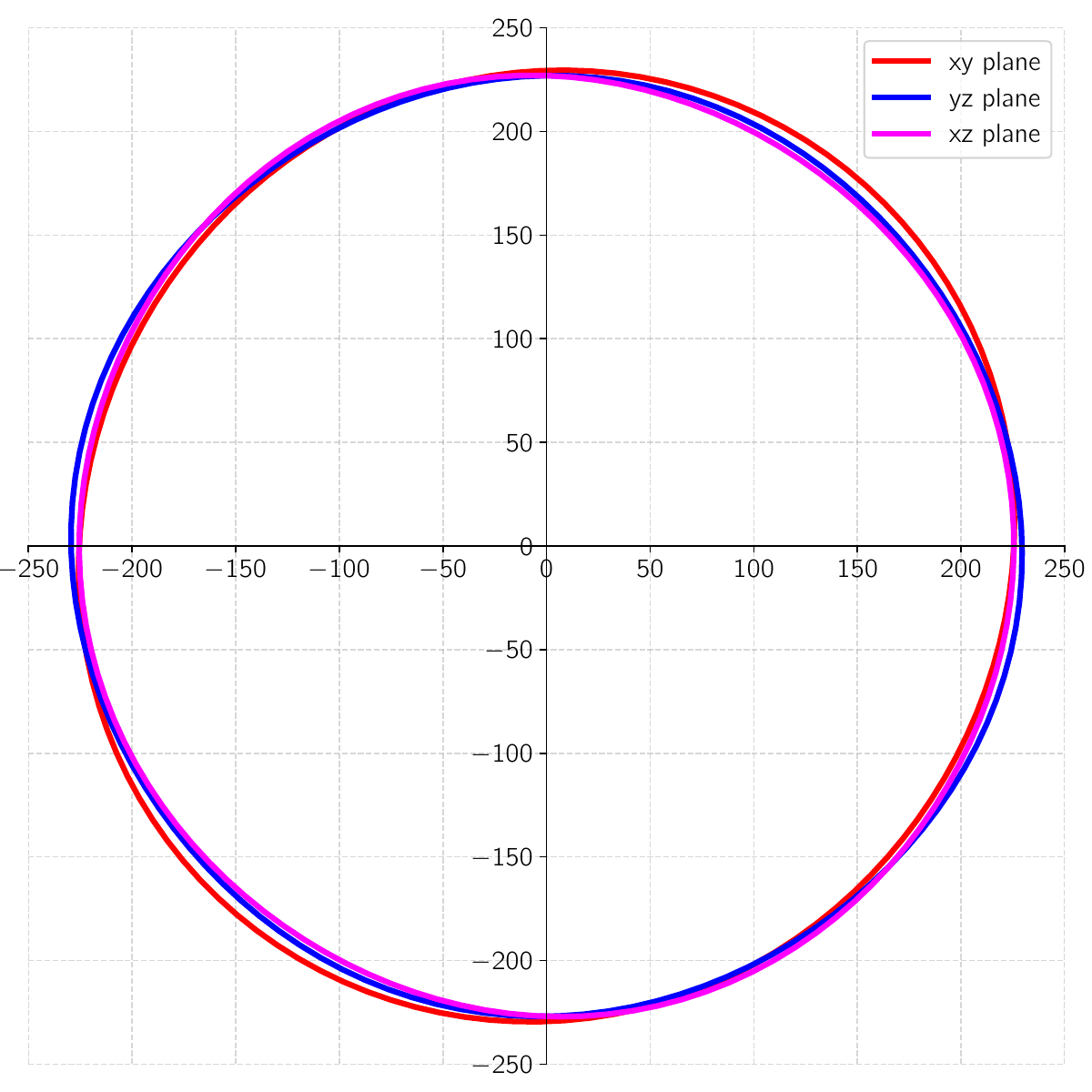}}}
  \caption{Directional Young's modulus from 0\,K calculations plotted
    along all the directions from (a)~\textit{ab initio} with 112
    atoms, MTP-based MD simulation with (b)~112 atoms (c)~7000
    atoms, and (d)~7000 atoms with exiting ML-GAP
    potential~\cite{Milardovich2023-we}. This shows the elastic
    response becomes isotropic with the increase in system size.}
  \label{fig:youngs}
\end{figure*}

\subsection{Elastic properties}

After obtaining models yielding a good description of the structural
properties of a-Si$_3$N$_4$, we shifted our discussion to the elastic
properties. 
As before, we predicted the elastic response of the a-Si$_3$N$_4$ by
\textit{ab initio} with 112 atoms in the first place, and compare these with MTP-based predictions.
Fig~\ref{fig:youngs_dft} shows the directional Young's modulus in
different planes estimated from the \textit{ab initio}-calculated
$6 \times 6$ elastic tensor. 
As amorphous materials are isotropic, we expect that the elastic
tensor of a-Si$_3$N$_4$ would not have any direction dependence, that
is, by using Eq.~\eqref{eq:isotropic_response}, $A^T = 1$. 
However, the value estimated for the \textit{ab initio} derived
elastic tensor with 112 atoms is $A^T=1.08$. 
Despite the large anisotropy shown in the Fig~\ref{fig:youngs_dft},
$A^T$ value for \textit{ab initio} does not deviate largely from the
ideal isotropic value, \textit{i.e.}, $A^T=1$. 
This deviation is because the value of $A^T$ was calculated from
selected elastic constants (simpler formulation of $A^T$ in
Eq.~\eqref{eq:isotropic_response}), whereas the directional Young's modulus reflects the full $6 \times 6$ symmetrical.
We, therefore, conclude that $A^T$ is not a sensitive enough measure in our particular case and will assess the isotropicity solely based on the visual representation of Young's modulus.

Next, we used ELATE~\cite{Gaillac2016-ss} to calculate the value of Young's modulus in the Voigt averaging scheme of polycrystal as 226 GPa, whereas for the Reuss and Hill averaging schemes, they are 216 and 221 GPa, respectively. 
Corresponding bulk moduli are 146, 143, and 145 GPa for Voigt, Reuss, and Hill, respectively.
It is noteworthy, that both quantities exhibit a relatively small spread (below 2.5\%) between the different averaging schemes.
Additionally, we calculated the elastic constants for other four different `melt-quench' structures of 112 atomic cells by \textit{ab initio} only to observe the spread of the Hill averages to be actually large: 217, 227, 255, 242 GPa (16\%). 
Such a large uncertainty prevents any trustworthy predictions of the tensorial elastic properties of this amorphous material based on structures consisting of 112 atoms.

Importantly, we reach the same conclusion also using MTP with the model having only 112: the resulting directional Young's modulus is significantly anisotropic~(Fig.~\ref{fig:youngs_mlip_122}).
Therefore, we probe a larger system with 7000 atoms, accessible only to classical MD simulations and enabled through our fitted MTP.
The results are plotted in Fig.~\ref{fig:youngs_mlip}, clearly demonstrating that Young's modulus overlaps in different directions, thereby representing a truly isotropic elastic response.
In this case, the anisotropy index according to Eq.~\eqref{eq:isotropic_response} yields $A^T=1.0$, which shows the elastic tensor is isotropic.
The polycrystalline Young's and bulk moduli are 226\,GPa and 174\,GPa, respectively (identical values up to 1\,GPa from Voigt, Reuss, and Hill averaging methods).
This shows that the MTP-based elastic response of this large model is isotropic.

Finally, we used the recently published ML-GAP potential to check for the elastic response of the large model with 7000 atoms. 
Again, the calculated Young's modulus (Fig.~\ref{fig:youngs_gap}) is fairly isotropic yielding values of Young's modulus, 227\,GPa, and bulk modulus, 174\,GPa, in excellent agreement with values predicted by our MTP.

We also point out that while the polycrystal Hill's average of Young's modulus is reasonably close to the value obtained from the large-scale simulation, the small model severely underestimates the bulk modulus (145\,GPa vs 174\,GPa).
In summary, simulation employing the large system size is crucial for
adequately describing tensorial properties, \textit{e.g.}, elastic response, of amorphous systems.

\subsection{Variability of mechanical properties}

\begin{figure}[tb]
  \centering
  \subfloat[\centering Elastic Modulus vs. Density]{\label{fig:denisty-blukmodulus}
    {\includegraphics[width=\linewidth]{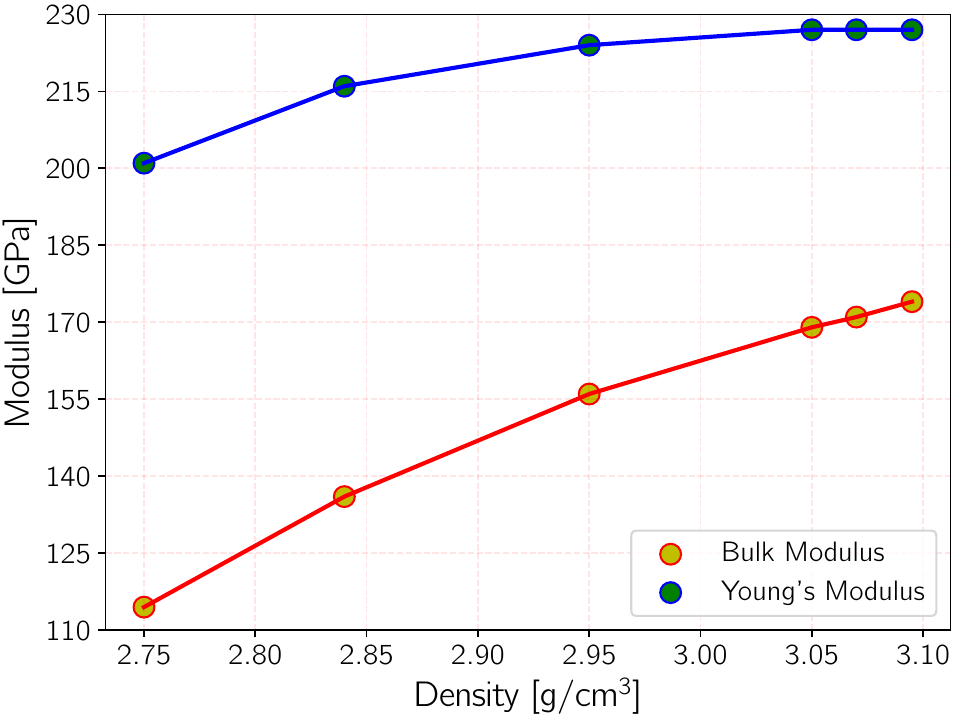}}}%
  \qquad
  \subfloat[\centering Elastic Modulus vs. Temperature]{\label{fig:T-bulkmodulus}
    {\includegraphics[width=\linewidth]{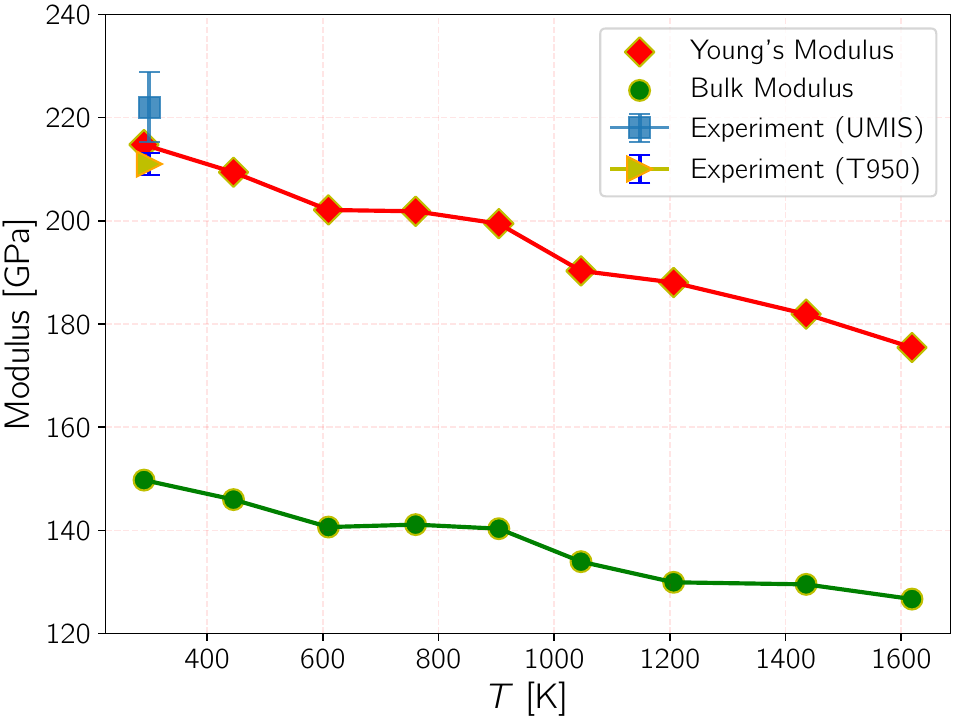}}}
  \caption{Estimation of mechanical properties in different
      conditions: (a)~Young's and bulk modulus for different mass
    densities. (b)~Variation of Young's and bulk modulus with
    temperature.}
  \label{fig:transferibility-mlip}
\end{figure}

Fig.~\ref{fig:transferibility-mlip} shows the calculated elastic properties in different conditions.
Firstly, we calculated bulk and Young's modulus for a series of mass densities (Fig.~\ref{fig:denisty-blukmodulus}).
Overall, the elastic moduli of the amorphous structure increase with the increase of mass density.
This is due to the decreasing of void spaces and, correspondingly, due to the increasing of the covalent bond density~\cite{Ito2014-xw}, which all together make the material
stiffer.

Secondly, the temperature dependence of the Young's and bulk moduli of a-Si$_3$N$_4$ is shown in Fig.~\ref{fig:T-bulkmodulus}. 
Both decrease with increasing temperature, \textit{ i.e.}, the material becomes more compliant with increasing temperature, in accordance with an intuitive expectation.

\subsection{Experiments, validation, and discussion}\label{sec:validation}

Apart from comparing the AIMD and MTP predictions, a series of
properties are also experimentally measured for further validation. 
The predicted density by \textit{ab initio} is $3.18\ \text{g}/\text{cm}^3$ (described in Sec.~\ref{sec:structural_a-SiN}).
This density is obtained as an average of a large number of small models optimized via AIMD followed by 0\,K structural optimization.
As mentioned in section~\ref{sec:structural_a-SiN}, the volumes directly approximated from the AIMD quenching runs yield larger volumes, which in turn means smaller densities.
These are scattered between $\approx2.65$\,g/cm$^3$ and $\approx 2.9$\,g/cm$^3$ for $T\to0$\,K, depending on the system size (cf. Fig.~\ref{fig:density}).
This latter data is informed about the system vibrational dynamics, similarly as the training data for our MTPs. 
It is, therefore, not surprising, that MTP predicted mass density is $2.94\ \text{g}/\text{cm}^3$ for both the quenching rates of $10^{13}$ and $10^{14}$\,K/s, whereas it slightly increases to $2.95\ \text{g}/\text{cm}^3$ for $10^{15}$\,K/s. 
Similarly, using the MTP to re-relax the structures of the small AIMD models yields 2.9\,g/cm$^3$, a value $\approx10\%$ smaller than the pure AIMD/DFT results.
All these values agree well with the previous findings~\cite{Milardovich2023-we,
  Vedula2012-kf, Aiyama1979-go}.
Our own measured value  $2.7\ \text{g}/\text{cm}^3$ by XRR is somewhat smaller, although well in the range of other experimental report~\cite{Aiyama1979-go}.

Despite the difference in the densities, the similarity in RDF (see
Fig.~\ref{fig:rdf}) and the structure factor (see
Fig.~\ref{fig:structure_factor}) among \textit{ab initio}, MTP, and
the experimentally measured data is astonishing.
The RDF and structure factor achieved through our simulations and XRD measurements, agree well with previously studied XRD-measured values~\cite{Aiyama1979-go}.
We, therefore, believe that the discrepancies in the \textit{ab initio} predicted mass densities stem from the genuinely underestimated volume.

Next, we compare the bond length and angle predictions from other
findings with us. 
\citet{Vedula2012-kf} reported the Si-N bond length for a series of mass
densities which are $(1.757 \pm 0.008)\,\uu{\AA}$, $(1.754 \pm
0.009)\,\uu{\AA}$ for mass density 2.9\,g/cm$^3$ and 3.1\,g/cm$^3$,
respectively, whereas respective N-Si-N bond angles are reported as
$(111.6\pm3.7)^{\circ}$ and $(110.3\pm2.4)^{\circ}$. 
Both the bond length and angles are agreed with our findings. 
The Si-N bond length from GAP ML potentials has been reported as 1.79\,$\uu{\AA}$ and the bond angle, $109.6^{\circ}$~\cite{Milardovich2023-we}, agrees well with our predictions.

Finally, we compare predicted elastic constants with those
experimentally measured by nanoindentation. 
For a fair comparison, we refer to our simulated room temperature
Young's modulus as the experimental measurement, was performed at room temperature. 
The calculated value at $300\,K$ is 220\,GPa (cf. Fig.~\ref{fig:T-bulkmodulus}). 
The Young's modulus value of $(222\pm7)$\,GPa, the UMIS nanoindenter measured, whereas the Hysitron T950 nanoindenter yielded a value of $(211\pm2)$\,GPa. 
Assuming a proper calibration of both instruments, the difference can be explained primarily by the inhomogeneity of the specimen, its surface roughness, and different load ranges (the sensitivity to surface imperfections is higher at lower loads). 
In comparison with previous reports from the sputtered deposited sample, where Young's modulus from nanoindentation is reported as $201\pm7$\,GPa for samples grown at RT and $210\pm7$ for growth at 850\,$^{\circ}$C~\cite{Vila2003-oe}, are in excellent agreement with our measurements.
In contrast to that \citet{Khan2004-ob} reported Young's modulus of $280\pm30$\,GPa, for density 3.1\,g/cm$^3$ in a CVD grown sample.
We ascribe the mismatch of this experimental value to the different sample preparation methods. 

Regarding theoretical values, \citet{Lehmann2001-ld} reported Young's modulus $237\pm54$ for the
mass density of 3.4\,g/cm$^3$ by molecular dynamics simulation. 
However, they used a higher content of Si than what corresponds to Si$_3$N$_4$, hence making any comparison to our predictions difficult.
\citet{Vedula2012-kf} reported a DFT value of bulk modulus of $169\pm3$\,GPa for mass density of 2.9\,g/cm$^3$, which is well comparable with our MTP prediction of 172\,GPa.

\section{Conclusions}

We have shown that moment tensor-based machine-learning interatomic
potential can accurately predict not only the structural properties of
amorphous materials, but also the tensorial properties, as
demonstrated by the elasticity calculations. 
Particularly, the latter was not possible using \textit{ab initio}
techniques considering the stringent limitations in terms of
model size \textit{i.e.}, up to a few hundred atoms. 
In the present work, we overcome this by a multi-method approach,
where we firstly trained MTP to our \textit{ab initio} data and
subsequently employed an MTP-based MD structural model of
a-Si$_3$N$_4$ containing 7000 atoms. 
The thus predicted structural as well as mechanical properties agree with experimentally measured values. 
These ﬁndings will have implications for future research on disordered
and amorphous materials, opening the door for quantitatively accurate
atomistic modeling with direct links to experiments for a-SiN$_x$ and
beyond. 

The main conclusions can be summarized as follows:
\begin{itemize}
\item MTPs can be used to model of amorphous materials accurately, provided there is sufficient
  training data sets either directly chosen or generated from active learning. 
\item A larger system size (beyond \textit{ab initio} capabilities) is
  necessary for precise calculation of tensorial elastic properties of amorphous materials.
\item By increasing the system size, one can design the appropriate
  ensemble or local atomic environments and corresponding short-range
  order interactions to model the disordered materials
  accurately. 
\item The variation of mechanical properties of a-Si$_3$N$_4$ is
  strongly mass density and deposition parameter dependent. Our
  predicted mechanical properties were found to be 220\,GPa at room
  temperature, which is in good agreement with the \textit{ab initio}, experimental, and other findings.
\item Regarding \textit{ab initio} only predictions, however, we note that (i) elastic response is strongly anisotropic and hence the above-mentioned agreement is rather accidental, and (ii) the equilibrium 0\,K density is overestimated due to not properly sampling the canonical ensemble of various structural motifs.
\end{itemize}

\section*{Acknowledgments}
GKN, RD, JT, PN and DH highly acknowledge the financial support
through the joint project of the Austrian Science Fund (FWF, project
number I~4059-N36) and the Czech Science Foundation (project number
19-29679L). P.S. would like to thank the Alexander von Humboldt Foundation for their support through the Alexander von Humboldt Postdoctoral Fellow- ship Program.
We acknowledge DESY (Hamburg, Germany), a member of the Helmholtz Association
HGF, is used for the provision of experimental facilities. Parts of this research were carried out at PETRA III, and we would like to thank Norbert Schell and Emad Maawad for their assistance in using beamline P07, operated by Helmholtz Zentrum Hereon. Beamtime was allocated for proposal I-20210616. The computational results presented have been partially achieved using the Vienna Scientific Cluster
(VSC).

\bibliographystyle{unsrtnat}

\bibliography{a-SiN_elasticity}

\end{document}